\documentstyle[preprint,revtex]{aps}

\begin{document}
\draft
\preprint{UM-P-92/28}
\preprint{OZ-92/08}
\begin{title}
Systematic Study of Fermion Masses and Mixing \\
Angles in Horizontal SU(2) Gauge Theory
\end{title}
\author{D. S. Shaw and R. R. Volkas}
\begin{instit}
Research Centre for High Energy Physics, School of Physics,\\
University of Melbourne, Parkville 3052, Australia
\end{instit}

\begin{abstract}
	Despite its great success in explaining the basic interactions
of nature, the standard model suffers from an inability to explain the
observed masses of the fundamental particles and the weak mixing angles
between them. We shall survey a set of possible extensions to the
standard model, employing an SU(2) ``horizontal'' gauge symmetry
between the particle generations, to
see what light they can shed on this problem.
\end{abstract}

\section{Introduction}

	One of the most important unresolved problems in particle
physics is the origin of the fermion mass spectrum and the mixing
angles between the quark fields. One avenue of investigation into this
problem is that of horizontal symmetry. This idea proposes a new
symmetry to exist between the different generations of particles.
Various symmetries have been investigated in the literature, both
discrete \cite{GEck,BG&S,KB&XGH} and continuous. Possible continuous
symmetries to look at are $U(1)$ \cite{DK&W1,DK&W2,AD&KW}, $SU(2)$
\cite{TM&TY,JChak,CLOng1,CLOng2,FW&AZ,KB&DC,MTY&WWW,FJL&V,KB&RM}
and $SU(3)$ \cite{TYan1,EP&GZ1,GZoup,EP&GZ2}
symmetries. While there has been considerable work on $SU(2)$
horizontal symmetry models, such investigations have been based on
only a few possible sets of fermion and Higgs boson assignments.

	In this paper we shall survey all the possible particle
assignments under $SU(2)$ looking at the predictions made by each
model for the tree level mass spectrum and the tree level
Kobayashi--Maskawa (KM) matrix. The gauge group is thus taken to be
$SU(3)_{C}\otimes SU(2)_{H}\otimes SU(2)_{L}\otimes U(1)_{Y}$, where
$SU(3)_{C}\otimes SU(2)_{L}\otimes U(1)_{Y}$ is the standard model
gauge group and $SU(2)_{H}$ is the horizontal gauge group.

	Before investigating this extended model, we shall take a brief
look at the qualitative features of the observed fermion mass and
mixing angle spectrum, to gain an idea of the sort of results we wish
to see coming from the models incorporating horizontal symmetry.

	Table \ref{masstab} displays the masses of the fundamental
charged fermions (neutrinos are taken to be massless in the first
part of this report --- bounds on the neutrino masses will be given in
Sec.\ \ref{addedRHNs}). The first and most obvious thing to note about
the mass spectrum is the progressive increase in mass with each
generation.
One way for this to come about, and the view that is taken in this
paper, is for the higher generation masses to result from first-order,
or tree-level, terms in the lagrangian.
In this paper, these tree-level interactions are taken
to be Yukawa couplings between fermion fields and Higgs multiplets.
The horizontal symmetry is spontaneously broken, with neutral Higgs
components acquiring non-zero vacuum expectation values (VEVs), and
through the tree-level Yukawa interactions with the particle
fields, this generates the
larger masses and mixing angles. The earlier generation masses are then
presumed to result from radiative corrections at the 1--loop level and
higher. To generate these radiative corrections will require further
extensions to the horizontal model.
{\em What we hope to see in the models
investigated in this report then, are tree-level mass spectra that
show masses for the heavier fermions and larger mixing angles, while
leaving the lighter fermion masses and smaller mixing angles to be
generated through higher order interactions.}

	Also notable about the mass spectrum is an approximate
hierarchy between the fermion sectors, with each charge
$\frac{2}{3}$ quark (the ``up'' quarks) being heavier in general than
the charge $-\frac{1}{3}$ quark (the ``down'' quark) of the same
generation, which in turn is heavier than the charged lepton of that
generation. The exception to this general trend is the up quark, whose
mass is a little lighter than the down quark mass. Hence, we might
expect to see tree-level masses for the
``up'' quarks only, or maybe the ``down'' quarks as well, while the
charged leptons (at least the lighter ones) would probably only gain
masses through higher-order corrections, which would be expected to be
naturally smaller.

	There is also the matter of the mixing angles between the two
quark sectors. Eq.\ (\ref{kmmatrix}) displays the experimentally
determined values for the entries in the Kobayashi--Maskawa (KM)
matrix \cite{PDG1}. To a first approximation, the KM matrix can be
taken as the identity matrix, so
following the above prescription,
one would hope to find this appearing at the tree level. Alternatively
would be a model that also generates the Cabibbo
angle at tree level, as this angle is of a comparable size to the
diagonal entries. The other two mixing angles are, however, relatively
small, so that while it may be feasible to generate the mixing angle
between the second and third generations at tree level (although this
would require some accurate cancellations between the parameters of
the model to keep this value small), it would be very improbable that
a model would naturally generate a tree level value for the mixing
angle between the first and third generations without also generating
one or both of the other angles.

\begin{equation}
U_{KM} = \left( \begin{array}{ccc}
0.9747-0.9759 & 0.218-0.224 & 0.001-0.007 \\
0.218-0.224 & 0.9734-0.9752 & 0.030-0.058 \\
0.003-0.019 & 0.029-0.058 & 0.9983-0.9996
\end{array} \right)
\label{kmmatrix} \end{equation}

	As will be seen, many of the models give quite reasonable mass
spectra, but they often have trouble with these mixing angle
hierarchy problems. Six models, however, do show
a fair amount of promise.

	The plan of the remainder of this paper is as follows:
Sec.\ \ref{models+eg} tables the possible representations
under $SU(2)_{H}$ for the various fermion fields
and all possible particle assignments consistent with anomaly
cancellation in the absence of right-handed neutrinos are determined.
A sample analysis of one model is given.
Sec.\ \ref{noRHNresults} presents an analysis of the results for the
various possible Higgs multiplets in each model (while it is
certainly feasible to combine various of the possible Higgs multiplets
within a model, in this report the results for each Higgs multiplet
within a given model will be looked at individually, with only
occasional comments on combining multiplets). In Sec.\ \ref{addedRHNs},
the effects of adding in right-handed neutrinos are investigated.
In Sec.\ \ref{extentions}, limits on horizontal gauge boson masses and
problems associated with horizontal symmetries are discussed and one
method for resolving some of these problems is presented. Sec.\
\ref{conclusion} gives a summary of important results.

\section{Anomaly-Free $SU(2)_{H}$ Spectra in the Absence \\
of Right-Handed Neutrinos}
\label{models+eg}

	The number of generations of particles is taken to be three.
This is in accord with data on the Z width from the LEP facility
\cite{LEP} which places the number of light neutrinos at
\begin{equation} N_{\nu} = 3.00 \pm 0.05 \end{equation}
While this does not exclude higher generations, the neutrinos in
these generations would be very massive and initially we will be
considering only models with massless neutrinos (no right-handed
neutrinos). Later, the possibility of including right-handed neutrinos
will be looked at, although the number of generations will still be
taken to be three.
	With three generations, each of the fermion fields $l_{\rm L},
e_{\rm R}, q_{\rm L}, u_{\rm R}$ and  $d_{\rm R}$ may transform under
$SU(2)_{H}$ as either a triplet, a doublet and a singlet or as three
singlets. To decide which of these combinations of representations are
reasonable, global and gauge anomaly cancellation is imposed.
Gauge anomaly cancellation is satisfied if the
$[\rm{SU}(2)_{H}]^{2}\rm{U}(1)_{Y}$ anomaly cancels out between the
lepton and quark fields, while global anomaly cancellation requires
that there be an even number of $\rm{SU}(2)_{H}$ doublets. Tables
\ref{leptontab} and \ref{quarktab} show the values for
these anomalies for the various possible ways of assigning
representations to the particle types.

	Comparing the values from these tables, fourteen theories
satisfying gauge and global anomaly cancellation appear. They are:
\begin{equation}
\begin{tabular}{ccc}
 L1 + Q1, & $\;\;\;\;\;\;$ & L1 + Q8, \\
 L2 + Q6, & & L3 + Q4, \\
 L5 + Q12, & & L5 + Q13, \\
 L6 + Q10, & & L6 + Q11, \\
 L7 + Q18, & & L8 + Q16, \\
 L1 + Q21, & & L9 + Q1, \\
 L9 + Q8, & & L9 + Q21. \\
\end{tabular} \label{noRHNmodels}
\end{equation}

	The last of these models is just the standard model, and
therefore of little interest to the present endeavour. For each
of the other models, the Yukawa couplings between
the fermions and Higgs particles will be
found, and the possible Higgs multiplets for that model determined.
{}From there, the
pattern of masses and mixing angles will be found. For the
sake of brevity, only one model is worked out in any detail here, with
the next section providing a table of results, and
a discussion of the features of the models.

	By way of illustration, we will now explicitly analyse the
\underline{L8 + Q16} model. We will first determine which Higgs
multiplets can couple to the quark and lepton bilinears through Yukawa
coupling terms. We will then take each possible Higgs multiplet in
isolation, and determine what fermion mass and mixing angle spectrum
results after spontaneous symmetry breaking.
The fermion assignments under
$\rm{SU}(3)_{C}\otimes\rm{SU}(2)_{H}\otimes
\rm{SU}(2)_{L}\otimes\rm{U}(1)_{Y}$
for this model are found to be:
\begin{equation}
\begin{array}{ll}
l_{iL} \sim (1,1,2)(-1), i=1,2,3 , & \\
e_{R} \sim (1,2,1)(-2), & e_{3R} \sim (1,1,1)(-2), \\
q_{L} \sim (3,2,2)(\frac{1}{3}), & q_{3L} \sim
(3,1,2)(\frac{1}{3}), \\
u_{R} \sim (3,2,1)(\frac{4}{3}), & u_{3R} \sim
(3,1,1)(\frac{4}{3}), \\
d_{iR} \sim (3,1,1)(-\frac{2}{3}), i=1,2,3 . &
\end{array}
\end{equation}
This gives the Higgs-fermion interactions
\begin{eqnarray}
{\cal L}_{\rm Yukawa}^{\rm H-f} & = & \lambda_{1i}\,\overline{l}_{iL}\,
e_{R}\,\chi^{C} + \lambda_{2i}\,\overline{l}_{iL}\,e_{3R}\,\phi^{C} +
\lambda_{3}\,\overline{q}_{L}\,u_{R}\,\phi \nonumber \\
 & & + \lambda_{4}\,\overline{q}_{L}\,u_{R}\,\Delta
+ \lambda_{5}\,\overline{q}_{L}\,u_{3R}\,\chi +
\lambda_{6i}\,\overline{q}_{L}\,d_{iR}\,\chi^{C} \nonumber \\
 & & + \lambda_{7}\,\overline{q}_{3L}\,u_{R}\,\chi +
\lambda_{8}\,\overline{q}_{3L}\,u_{3R}\,\phi +
\lambda_{9i}\,\overline{q}_{3L}\,d_{iR}\,\phi^{C} + {\rm H.c. ,}
\end{eqnarray}
where the Higgs multiplets are
\begin{equation}
\begin{array}{l}
\phi \sim (1,1,2)(-1), \\
\chi \sim (1,2,2)(-1), \\
\Delta \sim (1,3,2)(-1),
\end{array}
\end{equation}
and $\phi^{C}$ and $\chi^{C}$ are the charge conjugates of $\phi$ and
$\chi$ respectively --- $\phi^{C} \equiv i\tau_{2}\phi^{*}$ where
$\tau_{2}$ is the usual Pauli matrix operating in $SU(2)_{L}$ space
and $\chi^{C}$ results from similar transformations in both
$SU(2)_{L}$ space and $SU(2)_{H}$ space.

	We look at each of these three Higgs possibilities in
isolation. First, consider the terms involving $\phi$. This Higgs
multiplet has one neutral component, which gains a vacuum expectation
value (VEV), spontaneously breaking the gauge symmetry. Setting
\mbox{$\langle\phi^{0}\rangle = v$} gives the mass terms
\begin{equation} {\cal L}_{\rm Mass}^{\rm fermion} =
\overline{U}'_{L}M^{U}U'_{R} + \overline{D}'_{L}M^{D}D'_{R} +
\overline{E}'_{L}M^{E}E'_{R} + {\rm H.c.} \end{equation}
where $U' = (u',c',t')^{T}, D' = (d',s',b')^{T}, E' =
(e',\mu',\tau')^{T}$ (primes are used to denote that these are weak
eigenstates, not mass eigenstates). Also,
\begin{equation} M^{U} = \left( \begin{array}{ccc}
\lambda_{3}v & 0 & 0 \\
0 & \lambda_{3}v & 0 \\
0 & 0 & \lambda_{8}v
\end{array} \right) = D^{U}, \end{equation}
\begin{equation} M^{D} = \left( \begin{array}{ccc}
0 & 0 & 0 \\
0 & 0 & 0 \\
\lambda_{91}v\; & \lambda_{92}v\; & \lambda_{93}v
\end{array} \right) , \end{equation}
\begin{equation} M^{E} = \left( \begin{array}{ccc}
0\; & 0\; & \lambda_{21}v \\
0\; & 0\; & \lambda_{22}v \\
0\; & 0\; & \lambda_{23}v
\end{array} \right) , \end{equation}
where $M^{U}, M^{D}, M^{E}$ are the (undiagonalised) mass matrices for
the charge $\frac{2}{3}$ quarks, charge $-\frac{1}{3}$ quarks and the
charge $-1$ leptons respectively, and $D^{U}, D^{D}, D^{E}$ are the
corresponding diagonalised (mass eigenstate) matrices. The mass
eigenstate fields are of the form
\begin{eqnarray}
U'_{L} = A_{L}U_{L}, & U'_{R} = A_{R}U_{R}, & D'_{L} = B_{L}D_{L},
\nonumber \\
D'_{R} = B_{R}D_{R}, & E'_{L} = C_{L}E_{L}, & E'_{R} = C_{R}E_{R},
\end{eqnarray}
where $A_{L,R},B_{L,R},C_{L,R}$ are unitary matrices, and
\begin{equation} A_{L}^{\dagger}M^{U}A_{R} = D^{U},\;
B_{L}^{\dagger}M^{D}B_{R} = D^{D},\; C_{L}^{\dagger}M^{E}C_{R} = D^{E}.
\end{equation}
Clearly $A_{L} = A_{R} = I_{3\times3}$. It turns out that the simplest
choice for $B_{L}$ is $I_{3\times3}$
as well. Thus the KM matrix, given by $U_{KM} = A_{L}^{\dagger}B_{L}$,
is simply the identity matrix at tree level.
$D^{D}$ and $D^{E}$ are found to be:
\begin{equation} D^{D} =
(\lambda_{91}^{2} + \lambda_{92}^{2} +
\lambda_{93}^{2})^{\frac{1}{2}}v
\left( \begin{array}{ccc}
0\; & 0\; & 0 \\
0\; & 0\; & 0 \\
0\; & 0\; & 1
\end{array} \right) , \end{equation}
\begin{equation} D^{E} = (\lambda_{21}^{2} + \lambda_{22}^{2} +
\lambda_{23}^{2})^{\frac{1}{2}}v
\left( \begin{array}{ccc}
0\; & 0\; & 0 \\
0\; & 0\; & 0 \\
0\; & 0\; & 1
\end{array} \right) . \end{equation}
So the tree level results for this model are:

\begin{equation} \begin{array}{l}
{\rm Massive:\ u,c,t\ and\ b\ quarks\ and\ the\ lepton\ } \tau\ . \\
{\rm Massless:\ d\ and\ s\ quarks\ and\ the\ leptons\ e\ and\ }
\mu\ . \\
{\rm KM\ matrix:\ }U_{KM} = I_{3\times3}\ .
\end{array} \end{equation}

These tree-level results are quite good, with the exception of the
up-quark sector. Here we have either
\begin{equation}
m_{u} = m_{c}$ \qquad {\rm or} \qquad $m_{c} = m_{t}.
\end{equation}
Observationally, $m_{t} \gg m_{c} \gg m_{u}$. In order to satisfy this,
one must generate radiative corrections of similar magnitude to the
tree level results. If, for example, we take $m_{c} = m_{u}$ at tree
level, we need to generate radiative corrections of the order of the
(tree-level) charm quark mass, splitting the masses of the up and
charm quarks, and balanced so as to cancel the up quark mass almost
exactly (in comparison with the magnitude of the charm quark mass).
While this is not impossible, it seems unlikely that such precise
balancing would occur naturally.\vspace{0.5cm}

	We now move on to the next Higgs multiplet, $\chi \sim
\underline{2}$ (under SU$(2)_{H}$), which transforms according to
\begin{equation}
\chi \rightarrow U_{L} \chi U_{H}^{T},
\end{equation}
where $U_{L}$ represents a (unitary) rotation in $SU(2)_{L}$ space,
and $U_{H}$ is a (unitary) rotation in the horizontal space.
In matrix notation,
\begin{equation} \chi = \left( \begin{array}{cc}
\chi_{1}^{0} & \chi_{2}^{0} \\
\chi_{1}^{-} & \chi_{2}^{-}
\end{array} \right). \end{equation}
Both neutral components will gain VEVs when the horizontal symmetry
is broken: \mbox{$\langle\chi_{1}^{0}\rangle = w_{1}$},
\mbox{$\langle\chi_{2}^{0}\rangle = w_{2}$}.
This leads to
\begin{equation} M^{U} = \left( \begin{array}{ccc}
0 & \lambda_{5}w_{2} & 0 \\
-\lambda_{7}w_{1}\; & 0 & \;\lambda_{7}w_{2} \\
0 & \lambda_{5}w_{1} & 0 \end{array} \right) , \end{equation}
\begin{equation} M^{D} = \left( \begin{array}{ccc}
-\lambda_{61}w_{1}\; & -\lambda_{62}w_{1}\; & -\lambda_{63}w_{1} \\
0 & 0 & 0 \\
\lambda_{61}w_{2}\; & \lambda_{62}w_{2}\; & \lambda_{63}w_{2}
\end{array} \right) , \end{equation}
\begin{equation} M^{E} = \left( \begin{array}{ccc}
\lambda_{11}w_{2}\; & 0\; & \lambda_{11}w_{1} \\
\lambda_{12}w_{2}\; & 0\; & \lambda_{12}w_{1} \\
\lambda_{13}w_{2}\; & 0\; & \lambda_{13}w_{1}
\end{array} \right) . \end{equation}
Diagonalising these matrices gives
\begin{equation} D^{U} = (w_{1}^{2} + w_{2}^{2})^{-\frac{1}{2}}
\left( \begin{array}{ccc}
0\; & 0\; & 0 \\
0\; & \lambda_{7}\; & 0 \\
0\; & 0\; & \lambda_{5}
\end{array} \right) , \end{equation}
\begin{equation} D^{D} = [(\lambda_{61}^{2} + \lambda_{62}^{2} +
\lambda_{63}^{2})(w_{1}^{2} + w_{2}^{2})]^{\frac{1}{2}}
\left( \begin{array}{ccc}
0\; & 0\; & 0 \\
0\; & 0\; & 0 \\
0\; & 0\; & 1
\end{array} \right) , \end{equation}
\begin{equation} D^{E} = [(\lambda_{11}^{2} + \lambda_{12}^{2} +
\lambda_{13}^{2})(w_{1}^{2} + w_{2}^{2})]^{\frac{1}{2}}
\left( \begin{array}{ccc}
0\; & 0\; & 0 \\
0\; & 0\; & 0 \\
0\; & 0\; & 1
\end{array} \right) , \end{equation}
along with
\begin{equation} A_{L} = (w_{1}^{2} + w_{2}^{2})^{-\frac{1}{2}}
\left( \begin{array}{ccc}
w_{1} & 0 & w_{2} \\
0 & (w_{1}^{2} + w_{2}^{2})^{\frac{1}{2}} & 0 \\
-w_{2} & 0 & w_{1} \end{array} \right) , \end{equation}
\begin{equation} B_{L} = (w_{1}^{2} + w_{2}^{2})^{-\frac{1}{2}}
\left( \begin{array}{ccc}
w_{2} & 0 & -w_{1} \\
0 & (w_{1}^{2} + w_{2}^{2})^{\frac{1}{2}} & 0 \\
w_{1} & 0 & w_{2} \end{array} \right) . \end{equation}
Hence the KM matrix is
\begin{equation} U_{KM} = \left( \begin{array}{ccc}
0\; & 0\; & -1 \\
0\; & 1\; & 0 \\
1\; & 0\; & 0 \end{array} \right) . \end{equation}

	In such cases, a second $\chi$ Higgs multiplet may be added
(see ref. \cite{FJL&V}), having the effect of altering $U_{KM}$ to
\begin{equation} U_{KM} = \left( \begin{array}{ccc}
\cos \theta\; & 0\; & \sin \theta \\
0 & 1\; & 0 \\
-\sin \theta\; & 0\; & \cos \theta
\end{array} \right) . \label{KMeqn} \end{equation}

 It would be quite impressive if the angle $\theta$ could be identified
with the Cabibbo angle, however the location of the non-zero entries in
the diagonalised matrices $D^{U}$ and $D^{D}$ precludes such an
interpretation. The bottom-right entry in $D^{D}$ is taken to be the
bottom quark mass, implying that $\theta$ represents mixing between the
third and first generations (where $m_{t}$ is taken to be
$\lambda_{5}$ and $m_{c}$ is taken to be $\lambda_{7}$, that is, the
bases between which the KM matrix in Eq.\ (\ref{KMeqn}) operates are
the standard bases (u,c,t) and (d,s,b) --- other assignments are
possible but lead to KM matrices with zero diagonal entries). Since
experimentally the KM matrix entries $K_{13}$ and $K_{31}$ are found to
be very small ($\sim 0.004$), this is not a good result. It is possible
that the parameters of the model could be made to cancel so exactly as
to yield such a small quantity for $\theta$ and also generate the much
larger Cabibbo angle through radiative corrections, but the aim is to
find models in which this happens naturally, rather than requiring the
parameters to be carefully set ``by hand.''

	Hence, while the mass spectrum for this model is excellent,
providing tree level
masses for t,c,b and $\tau$ (in line with the observed relations $m_{t}
> m_{c} \simeq m_{b} \simeq m_{\tau} >$ all other masses), the model
suffers from a severe hierarchy problem in the KM matrix.

	A lesser problem, but one that crops up frequently, is the
hierarchy between those masses that {\em are\/} generated at tree
level. In this case, the same term in the lagrangian has generated
(generally) different masses for the top and the charmed quark, but a
large difference between the values of the coefficients $\lambda_{5}$
and $\lambda_{7}$ would still be required to explain why the observed
split between the masses of these two particles is as great as it is
(approximately two orders of magnitude).\vspace{0.5cm}

	The last Higgs multiplet for the \underline{$L_{8} + Q_{16}$}
model is the \mbox{$\Delta \sim \underline{3}$}\  multiplet. This is a
symmetric tensor in $SU(2)_{H}$ space, $\Delta_{\alpha}^{ij}$, where
$\alpha = 1,2$ indexes the weak isospin of the components and
$i,j = 1,2$ are $SU(2)_{H}$ indices. The $\Delta$ Higgs transforms
according to
\begin{equation}
\Delta_{\alpha}^{ij} \rightarrow
\Delta_{\beta}^{kl} = (U_{L})_{\beta}^{\alpha}(U_{H})^{k}_{i}
(U_{H})^{l}_{j}\Delta_{\alpha}^{ij} .
\end{equation}
There are three neutral components,
$\Delta_{1}^{0},\; \Delta_{2}^{0}\;{\rm and } \Delta_{3}^{0}$:
\begin{equation} \Delta_{(\alpha = 1)}^{ij} = \left( \begin{array}{cc}
\Delta_{1}^{0} & \frac{\Delta_{2}^{0}}{\sqrt{2}} \\
\frac{\Delta_{2}^{0}}{\sqrt{2}} & \Delta_{3}^{0}
\end{array} \right). \end{equation}
These neutral components acquire
VEVs \mbox{$\langle\Delta_{1}^{0}\rangle = x_{1}$},
\mbox{$\langle\Delta_{2}^{0}\rangle = x_{2}$},
\mbox{$\langle\Delta_{3}^{0}\rangle = x_{3}$}. Only
the ``up'' quarks couple to this Higgs, hence the ``down'' quarks and
the charged leptons will remain massless at
tree level. Furthermore the KM matrix becomes irrelevant at tree level
as the massless ``down'' quark states may be freely rotated into
superpositions of each other, making the KM matrix elements unphysical.
In the ``up'' quark sector,
\begin{equation} M^{U} = \left( \begin{array}{ccc}
\frac{x_{2}}{\sqrt{2}} & -x_{1} & 0 \\
x_{3} & -\frac{x_{2}}{\sqrt{2}} & 0 \\
0 & 0 & 0 \end{array} \right) , \end{equation}
leading to two massive quarks (c and t) upon diagonalising.
This easily explains the fermion sector hierarchy, with the charge
$\frac{2}{3}$ quarks having greater masses than the other fermions of
their respective generations, leaving room also for the up and down
quarks to have comparable masses, but it seems a little strange to
require that radiative corrections give masses to the bottom quark and
the $\tau$ comparable to the charm quark mass (it is possible if the
radiative corrections are proportional to the top quark mass, but the
hierarchy between the top and charmed quarks is unexplained).

\section{Tree Level Results}
\label{noRHNresults}

	Having analysed one model in detail, we now turn to table
\ref{sumtab} which provides a summary of the results for all
of the models (excepting the SM) in Eq. (\ref{noRHNmodels}).

	As can be seen from this table, in many cases the mass
spectrum is found to give satisfactory results. Typically, the third
generation only will receive masses, leaving the other generations to
obtain their lesser masses through radiative corrections. This occurs
in the models $\underline{L_{7} + Q_{18}}, \underline{L_{6} + Q_{11}},
\underline{L_{5} + Q_{12}}, \underline{L_{3} + Q_{4}}$ (for the
$\Delta$ Higgs multiplet) and $\underline{L_{1} + Q_{1}}$ ($\Delta$
Higgs multiplet).

	In other cases some or all of the second generation particles
also gain tree-level masses (the $\chi$ Higgs multiplet in
$\underline{L_{8} + Q_{16}}$, $\Delta$ in $\underline{L_{6} + Q_{10}}$,
$\chi$ and $\zeta$ in $\underline{L_{3} + Q_{4}}$, $\chi$ and $\Delta$
in $\underline{L_{2} + Q_{6}}$, and $\Delta$ in
$\underline{L_{1} + Q_{8}}$), although a couple of these models yield
strange relations, not existent in the observed spectrum, which seem
unlikely to be resolved naturally by the radiative corrections (an
example of this is the \mbox{$\chi \sim \underline{2}$}\ Higgs
multiplet from the model $\underline{L_{3} + Q_{4}}$ in which the
second generation masses are found to be related to the third
generation masses by the equation $m_{2} = \frac{m_{3}}{\sqrt{2}}$ ---
a relation which needs to be strongly broken by higher-order effects
in order to match the observed
difference in masses between these two generations. The $\chi$ Higgs
multiplet from the model $\underline{L_{2} + Q_{6}}$ also suffers
from this defect).

	A further problem with such models is that, while the one
Yukawa coupling term may generate different masses for two or three of
the fermions in a given sector (for instance, to the charm and top
quarks), it is hard to accept that this will naturally explain the
order-of-magnitudes splitting between the masses of particles of
different generations.

	Another pattern occasionally emerging is for just one or two
sectors (for example, only the ``up'' quarks) to be given masses. If
the sectors that receive masses are the heavier ``up'' quarks ---
possibly along with the ``down'' quarks --- then this is quite
feasible, especially if radiative corrections then couple together
particles within each generation, so that the bottom quark and the
tauon receive masses by interactions with the top quark and so on for
the other generations. Such a pattern agrees with the observed
hierarchy between the sectors, however the observed relation
$m_{u} \alt m_{d}$ may prove troublesome in such a scheme.
Models that follow this pattern are $\underline{L_{8} + Q_{16}}$
($\Delta$ Higgs multiplet), $\underline{L_{6} + Q_{10}}$ ($\phi$ and
$\psi$), $\underline{L_{5} + Q_{13}}$ ($\phi$),
$\underline{L_{2} + Q_{6}}$ ($\phi$), $\underline{L_{1} + Q_{8}}$
($\chi$), $\underline{L_{9} + Q_{8}}$ ($\chi$ and $\Delta$) and
$\underline{L_{9} + Q_{1}}$ (for the $\Delta$ and $\psi$ Higgs
multiplets).

	There are also a few models generating masses for all the
particles. Most, such as the \mbox{$\phi \sim \underline{1}$}\ Higgs
multiplet possibility from the model $\underline{L_{1} + Q_{8}}$, that
have equal masses (here, the leptons all have the same mass while two
charge $\frac{2}{3}$ quarks share one mass and two charge
$-\frac{1}{3}$ quarks share another) seem a little unlikely (other
models in this category are $\underline{L_{1} + Q_{1}}$,
$\underline{L_{9} + Q_{8}}$, $\underline{L_{9} + Q_{1}}$ and
$\underline{L_{1} + Q_{21}}$ --- looking at the $\phi$ Higgs
multiplet in each case). The one case that avoids these dubious
equalities, the $\psi$ Higgs transforming as a
$\underline{5}$ under $SU(2)_{H}$ in the $\underline{L_{1} + Q_{1}}$
model, may provide some reasonable
relations reducing the arbitrariness of the mass spectrum and the KM
matrix, but the algebra can be extremely dense and this case is of
limited interest. This model has, however,
received some attention from Wilczek and Zee \cite{FW&AZ} among others
\cite{KB&DC,MTY&WWW}. Their work focuses on a model combining the
$\Delta$ and $\psi$ multiplets from $\underline{L_{1} + Q_{1}}$, in
which the form of the Higgs potentials reduces the number of non-zero
components in the Higgs multiplets. The result is a model that
generates tree level masses for the second and third generations,
along with $U_{KM} = I_{3\times3}$.

	Although many of the models produce satisfactory results for
the mass spectrum of the particles, the KM matrix is often not so
good. For those cases where one of the quark sectors remains totally
free of any masses at tree level, then the KM matrix entries will be
unphysical (as is the case for the $\Delta$ Higgs possibility from
$\underline{L_{8} + Q_{16}}$, looked at in Sec.\ \ref{noRHNresults}).
Nothing further can be said about these models until the nature of the
radiative corrections is determined.

	An occasional model produces a KM matrix such as
\begin{equation} U_{KM} = \left( \begin{array}{ccc}
\;\;0\;\; & z\;\; & z\;\; \\
\;\;0\;\; & z\;\; & z\;\; \\
\;\;z\;\; & 0\;\; & 0\;\; \end{array} \right) , \end{equation}
where ``$z$'' represents a non-zero entry. These are tabulated as
``zero diagonal entries'', and clearly differ from the observed form
for this matrix.

	Many models produced finite values for some or all of the
off-diagonal entries in the KM matrix. Where the first and second
generation members of one (or both) quark sector(s) have zero mass at
tree level, the corresponding 2x2 sub-matrix in the KM matrix is
unphysical, and these entries can be eliminated. Often this leaves a
non-zero result for the angle(s) mixing the second and third
generation quarks and/or the first and third generation quarks. One
then has a hierarchy problem in explaining how these angles eventually
acquire their observed values. The problem is particularly severe
where the 1--3 generation mixing angle is non-zero, as this angle is
measured to have a value of about .004 that of the diagonal entries, a
hierarchy of the order of 1/200. The 2--3 generation mixing angle
hierarchy is less severe ($\sim 1/20$), and models generating this
angle at tree level are feasible --- though not very satisfactory.

	The most hopeful models are those for which the tree-level KM
matrix is simply the identity matrix, or those which generate a finite
value for only the 1--2 generation mixing angle (the Cabibbo
angle).\vspace{0.5cm}

	Of all the models, then, taking into account both the mass
spectra and the KM matrix results, six models look most
promising:\vspace{0.5cm}

	The first of these cases is the model
$\underline{L_{7} + Q_{18}}$ with the
\mbox{$\phi \sim \underline{1}$}\ Higgs multiplet. The
tree-level predictions for this model are:
\begin{equation} \begin{array}{l}
{\rm Massive:\ top\ and\ bottom\ quarks\ and\ the\ lepton\ } \tau\ . \\
{\rm Massless:\ up,\ down,\ charm\ and\ strange\ quarks\ and\ the\
leptons\ e\ and\ } \mu\ . \\
{\rm KM\ matrix:\ }U_{KM} = I_{3\times3}\ .
\end{array} \end{equation}
It is left to radiative corrections to assign smaller masses to the
initially massless first and second generation particles and to
provide values for the off-diagonal entries in the KM matrix. This
particular model has been looked at by Babu and Mohapatra \cite{KB&RM}.
In their paper, Babu and Mohapatra obtain radiative corrections for
the model by proposing the co-existence of the
\mbox{$\chi \sim \underline{2}$}\ Higgs
multiplet and also adding scalar fields of the type we describe in
Sec.\ \ref{extentions}. By assuming conservation of the tau lepton
number, and assigning $L_{\tau} = -1$ to the $\chi$ multiplet, the
neutral components of this multiplet are prevented from gaining VEVs,
thus leaving the mass matrices unchanged at the tree level, while
allowing for radiative corrections at
the 1--loop level and beyond.\vspace{0.5cm}

	The second model is $\underline{L_{3} + Q_{4}}$
with the \mbox{$\Delta \sim \underline{3}$}\ Higgs multiplet. The
predictions from this model are identical to those of the previous
model, namely tree level masses for the third generation particles
(t,b and $\tau$) and a tree-level KM matrix equal to the identity
matrix.\vspace{0.5cm}

	The next model, $\underline{L_{1} + Q_{8}}$ with the
\mbox{$\Delta \sim \underline{3}$}\ Higgs multiplet, generates masses
for most of the first and second generation particles:
\begin{equation} \begin{array}{l}
{\rm Massive:\ top,\ bottom,\ charm\ and\ strange\ quarks\ and\ the\
lepton\ } \tau\ . \\
{\rm Massless:\ up\ and\ down\ quarks\ and\ the\
leptons\ e\ and\ } \mu\ . \\
{\rm KM\ matrix:\ }U_{KM} = I_{3\times3}\ .
\end{array} \end{equation}
This approximates very well with the observed mass spectrum,
and again gives $U_{KM} = I_{3\times3}$ at tree level, although the
origin of the hierarchy between the second and third generations is
not clear from these tree-level results.\vspace{0.5cm}

	The $\underline{L_{1} + Q_{8}}$ model with the
\mbox{$\chi \sim \underline{2}$}\
Higgs multiplet proposes tree level masses for the four heaviest
quarks (t,b,c and s) leaving all the leptons massless. Also, if a
second $\chi$ Higgs multiplet is added in, this
model generates the Cabibbo angle at tree level. This is an
excellent result, except for the lack of a tree level mass for the
tauon. The masslessness of the leptons could possibly be fixed
by combining the $\chi$ Higgs multiplet with the
\mbox{$\psi \sim \underline{5}$}\ multiplet,
which generates (distinct) masses for all three leptons. If this is
done then the radiative corrections will need to be able to explain
how the initially massless first generation quarks (u and d) receive
masses generated at higher order which are of greater magnitude than
the (tree-level) electron mass.\vspace{0.5cm}

	Finally, two Higgs multiplets from the
$\underline{L_{9} + Q_{8}}$ model look promising. In this model, only
the quark fields transform in a non-trivial manner under the horizontal
symmetry, the leptons behaving just as they do in the SM. Both of the
Higgs multiplets generate masses for the second and third generation
quarks, leaving the leptons massless at tree level. The
\mbox{$\chi \sim \underline{2}$}\ Higgs
multiplet generates a tree level Cabibbo angle,
while the other Higgs multiplet, \mbox{$\Delta \sim \underline{3}$}\,
leaves the KM matrix equal to the identity matrix at tree level. Again,
these results are in good agreement with the observed mass spectrum,
but the masslessness of the tauon is a little
problematical.\vspace{0.5cm}

\section{Models Incorporating Right-Handed Neutrinos}
\label{addedRHNs}

	This section will look at the effect of including
right-handed neutrinos (RHNs) in the particle spectrum. Since these
neutrino fields are invariant under transformations of the SM gauge
group (although not necessarily under the horizontal gauge group),
there will be no contribution from these fields to the gauge anomaly.
The contribution of the RHN fields to the global anomaly will also be
zero if the neutrinos transform either as three singlets or as a
triplet under $SU(2)_{H}$ (``N1'' or ``N3'' respectively), and will be
one doublet if the fields transform as a singlet and a doublet
(``N2''). This leads to a total of 35 models for analysis:
\begin{equation}
\begin{tabular}{ccc}
L1 + Q1 + N1, & $\;\;\;\;\;\;$ & L1 + Q1 + N3, \\
L1 + Q8 + N1, & & L1 + Q8 + N3, \\
L2 + Q5 + N2, & & \\
L2 + Q6 + N1, & & L2 + Q6 + N3, \\
L3 + Q3 + N2, & & \\
L3 + Q4 + N1, & & L3 + Q4 + N3, \\
L4 + Q1 + N2, & & \\
L4 + Q8 + N2, & & \\
L5 + Q12 + N1, & & L5 + Q12 + N3, \\
L5 + Q13 + N1, & & L5 + Q13 + N3, \\
L6 + Q10 + N1, & & L6 + Q10 + N3, \\
L6 + Q11 + N1, & & L6 + Q11 + N3, \\
L7 + Q18 + N1, & & L7 + Q18 + N3, \\
L7 + Q19 + N2, & & \\
L8 + Q16 + N1, & & L8 + Q16 + N3, \\
L8 + Q17 + N2, & & \\
L1 + Q21 + N1, & & L1 + Q21 + N3, \\
L4 + Q21 + N2, & & \\
L9 + Q1 + N1, & & L9 + Q1 + N3, \\
L9 + Q8 + N1, & & L9 + Q8 + N3, \\
L9 + Q21 + N1, & & L9 + Q21 + N3.
\end{tabular}
\end{equation}

	Some new complications arise in the analysis of these models.
First, along with the KM matrix operating between the quark fields,
we now have a lepton mixing matrix, $U_{e\nu}$, operating between the
charged leptons and the neutrinos. Second, experiments to date have
credited the neutrinos with either zero, or very small, masses ---
typically orders of magnitude less than for the other particles in a
given generation. Table \ref{nu-bounds} shows the current bounds on the
neutrino masses resulting from accelerator and double-beta decay
measurements. This implies a disquietingly large disparity between
the values of the Yukawa coupling constants in the mass terms for the
neutrinos and the corresponding terms for the other fermion sectors.
One way of avoiding this disparity is to invoke the see-saw mechanism
\cite{GMRS,TYan2,RM&GS} which allows for comparable Dirac masses for
all the particles, but provides small observable Majorana masses for
the neutrinos.
	Table \ref{rhnsummary} displays the results for the above
models (with the exception of the model $\underline{L9 + Q21 + N1}$
which is the same as the SM with RHNs and no horizontal symmetry).

	Since the charged fermion sectors have been dealt with,
in studying these models all of the comments made in Sec.\
\ref{noRHNresults} will apply here also. Many of the cases can be
disregarded because they suffer from problems in the quark sectors or
the charged lepton sector. Others suffer from unfavourable KM matrix
results. Those models that show promise after this initial filtering
are then subject to analysis to see what they predict for the neutrino
sector.

	For several of these models, it will be seen that the RHNs do
not couple to some of the possible higgs multiplets. In these cases
the neutrinos remain massless at tree level, acquiring masses only ---
if at all --- through radiative corrections, providing a natural
reason for the small or zero masses observed for these particles.

	For other cases, investigation is hampered by the lack of data
concerning the precise masses of the neutrinos, and the values of the
mixing angles between the two lepton sectors. Leaving aside the
possibility that the neutrinos are, in fact, massless, the smallness
of their masses can be explained through the so-called see-saw
mechanism. Here, the neutrinos initially acquire Dirac masses of
comparable size to the masses of the other fermion sectors. The RHN
fields, however, also develop very large Majorana masses. The observed
neutrino masses are then those resulting from diagonalising the matrix
below:

\begin{equation} {\cal L}_{\rm mass} =
(\overline{\nu_{L}},\overline{(\nu_{R})^{C}})\left( \begin{array}{cc}
0 & m \\
m^{T} & M
\end{array} \right) \left( \begin{array}{c}
(\nu_{L})^{C} \\ \nu_{R}
\end{array} \right), \end{equation}

where $m$ is a three-by-three matrix of Dirac masses, and $M$ is the
corresponding matrix of Majorana masses. The eigenvalues for this
matrix are, for $M$ large, $M$ (assumed to be unobserved as yet) and
$m^{2}/M$. The latter value is the observed result, and can
clearly be very small for large values of $M$.
In general, for a similar spread of
values, the Dirac masses should have more impact on the (observable)
neutrino mass hierarchy as the latter depend on the difference of
the squares of the Dirac masses, while only depending on the (linear)
difference of the Majorana masses.

	It is worth noting that the Majorana terms in the
Lagrangian may arise from Yukawa couplings of the form
\begin{equation} {\cal L}_{\rm Yuk} =
\lambda\;\overline{(\nu_{R})^{C}}\;\sigma\;\nu_{R}
\end{equation}
where $\sigma$ is a scalar under the standard model gauge group, but
not necessarily under the horizontal gauge group. For example, if the
RHNs transform as a triplet under $SU(2)_{H}$, then $\sigma$ could be
either a $\underline{1}$, $\underline{3}$ or $\underline{5}$ under the
horizontal gauge group. Such scalar fields are ideal both for breaking
the horizontal symmetry at a high energy and for generating radiative
corrections, as described in Sec.\ \ref{extentions}. Furthermore, the
very high VEVs that these fields need to develop to fulfill this
function automatically assure that the Majorana masses will be
extremely high.

	In the table, the neutrinos have been labelled $\nu_{1}$,
$\nu_{2}$ and $\nu_{3}$, since it cannot be said what mixture of these
mass eigenstates corresponds to the weak eigenstates, $\nu_{e}$,
$\nu_{\mu}$ and $\nu_{\tau}$. This is, of course, because the nature
of the lepton mixing matrix is largely unknown. If the off-diagonal
entries of this matrix were large, however, then non-conservation of
the individual lepton numbers $L_{e}$, $L_{\mu}$ and $L_{\tau}$ should
be easy to observe. The lack of such observations --- particularly for
$\nu_{e} \leftrightarrow \nu_{\mu}$ exchanges --- places severe limits
on these values. This suggests that, at least, $\nu_{1}$ is the
principle component of the electron neutrino.

	In order to further investigate any of the cases above, it
becomes necessary to choose a particular model for the neutrino mass
spectrum. Here, we shall look at a paper by Caldwell and Langacker
\cite{DC&PL}. They take as their starting point the 17 keV neutrino
claimed to exist by Simpson and
others \cite{Simp,JS&AH,AH&JS,BS+,EBN,IZ,AH&NAJ}. By looking at the
effects of various constraints,
Caldwell and Langacker reached the following conclusions. Firstly, the
17 keV neutrino is assumed to be $\nu_{3}$, and is found to be the
dominant component of $\nu_{\tau}$. Secondly, $\nu_{\mu}$ is a heavy
Majorana neutrino, with a mass of either about 17 keV or in the range
170--270 keV.

	For the case that the masses of $\nu_{2}$ and $\nu_{3}$ are
both 17 keV, with the muon and tauon neutrinos being almost pure
mixtures of these two mass eigenstates, a further symmetry is possible.
Here, the mixing angles between the first and second and first and
third generation leptons are small and equal, while the second-third
generation angle, $\theta_{3}$, satisfies \[ \cos\theta_{3} =
\frac{1}{\sqrt{2}} = -\sin\theta_{3} .\]
This symmetry conserves $L_{e} - L_{\mu} + L_{\tau}$. Small deviations
from this symmetry are possible, but the exact symmetry would be
expected to show through at the tree level. One candidate model for
such a symmetry is the $L_{3} + Q_{4} + N_{3}$ model with the $\zeta
\sim \underline{4}$ higgs multiplet, which suffers from a
second-third generation mixing angle hierarchy problem in the KM
matrix, and does not explicitly equate the neutrino masses, but is
otherwise suitable (perhaps a further relation, maybe coming from a
more thorough analysis of the higgs potential, can restrict the model
further, forcing the neutrino masses to be equal and giving the mixing
angle its appropriate value). None of the other cases looks promising
for this first scenario.

	The second possibility found by Caldwell and Langacker has all
the lepton mixing angles small, with 17 keV for the mass of $\nu_{3}$,
and hence for $\nu_{\tau}$, and 170-270 keV for the mass of $\nu_{2}$,
which is the dominant component of $\nu_{\mu}$. It may be, in this
case, that the muon neutrino mass is generated at tree level (thus
engendering a name change in table \ref{rhnsummary}: $\nu_{3}
\leftrightarrow \nu_{2}$), while the 17 keV mass arises from radiative
corrections. Feasible models for generating this scenario are the
$L_{7} + Q_{18} + N_{1}$ model with the $\phi$ higgs, transforming as
a scalar under the horizontal symmetry; the $L_{4} + Q_{8} + N_{2}$
model with $\Delta \sim \underline{3}$ higgs multiplet (the $\chi$
Higgs multiplet in this model is investigated in Ref. \cite{FJL&V},
it provides tree level masses for the second and third generation,
including neutrinos, but generates non-zero lepton mixing angles at
tree level, which therefore may be quite large); the model
$L_{3} + Q_{4} + N_{1}$, with $\phi$ higgs (here, {\em all \/} the
fermion masses and mixing angles would need to be generated from
mixing with $\nu_{2}$ through radiative corrections. This seems a
little dubious, especially since one would expect the second
generation particles to receive greater masses than the third
generation. One way around this is to combine the $\phi$ Higgs with
the $\Delta$ Higgs, which provides masses for the charged third
generation particles t, b and $\tau$);
the $L_{3} + Q_{4} + N_{3}$ model with the $\Delta \sim
\underline{3}$ higgs multiplet; and the $L_{9} + Q_{8} + N_{3}$ model
with the $\Delta$ higgs multiplet (although the lack of a mass for the
tauon at tree level is a little odd).

\section{Other Aspects of $SU(2)_{H}$ Models}
\label{extentions}
\nobreak

	Having examined the tree level results for models based
on a horizontal $SU(2)$	gauge symmetry, some general points about
such models are in order. Firstly, these models will generate
flavour changing neutral currents (FCNCs) mediated by $SU(2)_{H}$
gauge bosons. FCNCs will also be generated in the Higgs sector. The
FCNCs resulting from Higgs couplings can be suppressed by giving
the Higgs bosons sufficiently large masses. Since the Higgs sectors we
have studied serve only as starting points for more realistic mass
generation schemes, we will not pursue any detailed Higgs
phenomenology in the present paper. Note, however, the generic feature
that the Higgs-induced FCNC processes will be weaker for the lighter
fermions compared with the heavier fermions.

	The gauge-boson generated FCNCs may be suppressed by breaking
the horizontal symmetry at a high energy scale, thus generating large
masses for the bosons. The following is a simplified,
order-of-magnitude examination of minimum gauge boson masses due to
constraints from the observed smallness of certain FCNC processes.
More precise calculations of the mass limits would be cumbersome and
of limited worth since either the minimum values for the gauge boson
masses are found to be beyond the energies of any present-day
operating or planned facility or other factors preclude the likely
observation of the bosons (an example of such a case is included
below). The horizontal gauge boson masses are assumed to be degenerate
and mixing with the Z boson is assumed to be very small. This
simplifies the discussion without seriously affecting the results. The
actual mechanism for generating the horizontal gauge boson masses is
outlined later in this section.

	It is found that many models will not support various of the
FCNC processes that might be considered. In particular, it is worth
noting that the tree-level horizontal contribution to the $K_{S}-K_{L}$
mass difference will cancel out in every model, even when a tree-level
Cabibbo angle is generated. The coupling here is of the form
\begin{equation}
\lambda \overline{f}_{L,R} U^{\dagger} T^{a} U \gamma^{\mu} f_{L,R}
H^{a}_{\mu}
\end{equation}
where $f_{L,R}$ is the mass eigenstate fermion
field, $U$ is the unitary matrix rotating the weak eigenstates of
the fermions into mass eigenstates,
$T^{a}$ are the generators of the horizontal symmetry
and $H^{a}$ are the associated horizontal bosons. If the same field
$f_{L,R}$ couples to each end of the horizontal boson propagator,
then it turns out that the angles (which exist because the Higgs
multiplets and the fermion fields transform non-trivially under
$SU(2)_{H}$) cancel out in the sum over $a$, resulting in a zero
contribution to the process.

	Hence, different processes will provide the best constraints
on the horizontal boson masses for different models. In this paper,
three different processes shall be looked at, chosen to exemplify the
main types of processes to which the horizontal models can contribute.

	First, the lepton-number violating decay
\begin{equation}
\tau^{-} \rightarrow \mu^{-}e^{+}\mu^{-}
\end{equation}
shall be looked at. This decay occurs, for example, in the
$\underline{L_{1} + Q_{8}}$ model with the Higgs multiplet and the
leptonic fermion fields
transforming as triplets under $SU(2)_{H}$. Fig. (\ref{taudecay})
shows a diagram of the decay for this model. For simplicity, the
Lorentz form of the interactions will be ignored, with the result
coming from a comparison with the branching ratio of a similar decay.
For example, in this instance, the branching ratio of the lepton-number
violating decay will be compared to that of the decay

\begin{equation}
\tau^{-} \rightarrow \mu^{-}\nu_{\tau}\overline{\nu}_{\mu}.
\end{equation}

	Thus, one arrives at the expression ($g_{H},\; M_{H},\; g_{W}\;
{\rm and\ } M_{W}$ represent respectively the horizontal coupling
constant and gauge boson mass, the weak coupling constant and the W
boson mass)
\begin{equation}
\frac{g_{H}^{2}}{M_{H}^{2}} \leq \frac{{\rm BR}(\tau \rightarrow \mu e
\mu) \times g_{W}^{2}}{{\rm BR}(\tau \rightarrow \mu \nu
\overline{\nu}) \times M_{W}^{2}}.
\end{equation}

	The experimental value for the SM decay is BR$(\tau
\rightarrow \mu \nu \overline{\nu}) = 0.178 \pm 0.004$ \cite{PDG2}
while the limit on the lepton-number violating decay is
BR$(\tau \rightarrow \mu^{-} e^{+} \mu^{-}) < 3.8 \times 10^{-5}$
\cite{PDG3}, leading to the limit
\begin{equation}
\frac{g_{H}^{2}}{M_{H}^{2}} \leq 2.18 \times 10^{-4} \times
\frac{g_{W}^{2}}{M_{W}^{2}}.
\end{equation}
In the case of the coupling constants being equal, the lower limit on
the horizontal gauge boson mass becomes (taking $M_{W} = 80.6$ GeV
\cite{PDG4})
\begin{equation}
M_{H} \agt 5 \; {\rm TeV.}
\end{equation}

	One of the most stringent limits on the gauge boson masses
comes from the semi-leptonic decay
\begin{equation}
K^{-} \rightarrow \pi^{-}e^{+}\mu^{-},
\end{equation}
which occurs, for example, in three of the six more promising models
discussed in Sec.\ \ref{noRHNresults}. In these models the first two
generations typically form doublet fields under the horizontal
symmetry. In the $\underline{L_{7} + Q_{18}}$ model with
\mbox{$\phi \sim \underline{1}$}\ Higgs multiplet, the decay
will proceed as shown in Fig. (\ref{kaondecay}). In this case, we
arrive at the approximate ratio

\begin{equation}
\frac{g_{H}^{2}}{M_{H}^{2}} \leq \frac{g_{W}^{2} \times
\sin^{2}{\theta_{c}} \times {\rm BR}(K^{-} \rightarrow \pi^{-} e^{+}
\mu^{-})}{2 \times {\rm BR} (K^{-} \rightarrow \pi^{0} \mu^{-}
\overline{\nu}_{\mu}) \times M_{W}^{2}},
\end{equation}
where $\theta_{c}$ is the Cabibbo angle.

	Observations to-date \cite{PDG5} put the rate of the FCNC
decay at
\begin{equation}
{\rm BR}(K^{-} \rightarrow \pi^{-} e^{+} \mu^{-}) < 2.1 \times
10^{-10},
\end{equation}
while the SM decay rate is \cite{PDG6}
\begin{equation}
{\rm BR}(K^{-} \rightarrow \pi^{0} \mu^{-} \overline{\nu}_{\mu}) =
0.0318 \pm .0008,
\end{equation}
leading to the bound
\begin{equation}
\frac{g_{H}^{2}}{M_{H}^{2}} \leq 1.69 \times 10^{-10} \times
\frac{g_{W}^{2}}{M_{W}^{2}},
\end{equation}
or, for equal-valued coupling constants, we get the extreme mass limit
\begin{equation}
M_{H} \agt 6 \times 10^{3} \; {\rm TeV.}
\end{equation}

	Finally, we take a look at the model
$\underline{L_{9} + Q_{8}}$ with \mbox{$\Delta \sim \underline{3}$}\
Higgs. This model proves interesting in that only the second and
third generation quark fields transform non-trivially under the
horizontal symmetry. Thus, the only
allowed FCNC processes must involve mixing between either the top and
charm quarks or the bottom and strange quarks. Thus the most extreme
limits will come from processes such as the decay of the $\Upsilon$
meson into a $\phi$ meson (Fig. (\ref{upsilondecay}) shows one
possible channel for this decay, with gluon emission from one of
the final-state quarks for kinematic balance). The limit from
decays such as this, though, will be very weak as little experimental
information is available on the particles involved and the
experimental uncertainties are comparatively large. While it is
possible, then, that the horizontal bosons could have fairly low
masses in such models, production and investigation of the bosons will
be hampered by the lack of tree-level interactions
between the horizontal bosons and the first-generation
fermion fields.

	Another concern of horizontal models of this kind is how to
generate the radiative corrections that are assumed to provide the
smaller masses and mixing angles in the models. What follows is a
brief description of one method for achieving this \cite{FJL&V,BSBal}.

	First, a scalar field ($\sigma$) is introduced that transforms
as a singlet under the SM gauge group, but not under $SU(2)_{H}$. This
is a neutral field and is assumed to pick up a large VEV, breaking the
horizontal symmetry at a high energy scale. As well as leading to
radiative corrections, this will generate the large masses for the
horizontal gauge bosons discussed above. As has
been noted in the previous section, such neutral scalars may appear
anyway in the Majorana neutrino terms in the lagrangian. There, the
large VEVs for these scalar fields leads to suitably large Majorana
masses and the see-saw mechanism then naturally explains the smallness
of observed neutrino masses.

	Further scalar fields are introduced ($\eta, \eta'$) which
are charged (thus they acquire no VEVs and will not affect the tree
level results) and which couple with the neutral scalar field,
inducing mass mixing between them. These latter fields will also
couple to the fermion fields, producing 1--loop diagrams such as
that shown in Fig. (\ref{oneloopdia}) that will provide small
corrections to the masses and mixing angles.

	An interesting feature of many models with masses generated
through radiative corrections is that they couple the up quark to the
strange and bottom quarks and couple the down quark to the charm and
top quarks through the higher-order corrections, thus providing a
natural explanation for the observation $m_{u} < m_{d}$.

\section{Conclusion}
\label{conclusion}

	An SU(2) gauge symmetry between the three generations of
fermions is one possibility for explaining the major features of the
observed fermion mass and mixing angle spectrum.

	If there are no RHN fields, then there are ten ways to assign
representations under this symmetry to the different fermion fields
such that gauge and global anomalies cancel. We have surveyed each of
these models to determine which possible Higgs multiplets lead to
promising tree level mass relations and KM matrices. A brief summary
of the results for each model is given in table \ref{sumtab}.

	In the case where RHN fields are also assumed to exist, a
further 35 models satisfying anomaly cancellation are possible. While
many of these models can be discarded as unlikely to represent nature
because of problems in the charged fermion sectors, analysis of the
predictions for neutrino masses is hindered by the lack of definite
information regarding these masses. We investigated the models
containing RHNs, taking as assumptions the existence of the claimed
17keV neutrino and following the study of Caldwell and Langacker into
the consequences of this assumption.

	For the future, there are various areas of work to follow up
on. Having found the tree level results for a given model, the next
step is to work out the details of generating radiative corrections,
for which a possible method is outlined in Sec.\ \ref{extentions}.
Also, further phenomenology associated with horizontal $SU(2)$
symmetry models, such as the Higgs potentials --- which may show how
to improve on some of the models by preventing some of the neutral
Higgs components from gaining VEVs --- may be investigated.

	We wish to thank Andrew Davies for some useful discussions.

\figure{Horizontal gauge boson induced, lepton number
violating decay.\label{taudecay}}

\figure{Horizontal gauge boson induced kaon decay.\label{kaondecay}}

\figure{Horizontal gauge boson induced Upsilon
decay. While gluon emission by one of the final-state quarks will
dominate, other emissions such as photon or Z boson emission are
also possible.\label{upsilondecay}}

\figure{A one-loop level diagram of the type that leads to
masses for the lighter fermions. Here, an initially massless
fermion (${\rm f}_{L}, {\rm f}_{R}$) is converted to and from a heavy
fermion (${\rm F}_{L}, {\rm F}_{R}$) through the emission of charged
scalars ($\eta, \eta'$).
These scalars couple via a neutral scalar coming out of the vacuum
($\langle \sigma \rangle$). Thus the light fermion will pick up a mass
which is dependent on the mass of the heavy fermion ($m_{F}$), the VEV
of the neutral scalar, the masses of the charged scalars and the
Yukawa coupling constants of the charged scalars.\label{oneloopdia}}

\begin{table}
\caption{\label{masstab} Observed Masses of Fundamental Particles
\protect\cite{PDG7,PDG8}}
\begin{tabular}{crcrcr}
$m_{\rm up}$ & $\sim 5.6 \pm 1.1$ MeV & $m_{\rm charm}$ & $\sim 1.35
\pm 0.05$ GeV & $m_{\rm top}$ & $>$ 89 GeV \\
$m_{\rm down}$ & $\sim 9.9 \pm 1.1$ MeV & $m_{\rm strange}$ & $\sim 199
\pm 33$ MeV & $m_{\rm bottom}$ & $\sim 5$ GeV \\
$m_{\rm elec}$ & 0.511 MeV & $m_{\rm muon}$ & 105.7 MeV &
$m_{\rm tauon}$ & $1784.1^{+2.7}_{-3.6}$ MeV \\
\end{tabular}
\end{table}

\begin{table}
\caption{\label{leptontab} Possible Lepton $SU(2)_{H}$ Representation
Assignments}
\begin{tabular}{ccccc}
 & $l_{L}$ & $e_{R}$ & Gauge Anomaly & Global Anomaly \\
 & & & Contribution & (\# of doublets) \\ \tableline
$L_{1}$ & \underline{3} & \underline{3} & 0 & 0 \\
$L_{2}$ & \underline{3} & $\underline{2}\oplus\underline{1}$ & $-6$ & 1
\\
$L_{3}$ & $\underline{2}\oplus\underline{1}$ & \underline{3} & 6 & 2 \\
$L_{4}$ & $\underline{2}\oplus\underline{1}$ &
$\underline{2}\oplus\underline{1}$ & 0 & 3 \\
$L_{5}$ & \underline{3} &
$\underline{1}\oplus\underline{1}\oplus\underline{1}$ & $-8$ & 0 \\
$L_{6}$ & $\underline{1}\oplus\underline{1}\oplus\underline{1}$ &
\underline{3} & 8 & 0 \\
$L_{7}$ & $\underline{2}\oplus\underline{1}$ &
$\underline{1}\oplus\underline{1}\oplus\underline{1}$ & $-2$ & 2 \\
$L_{8}$ & $\underline{1}\oplus\underline{1}\oplus\underline{1}$ &
$\underline{2}\oplus\underline{1}$ & 2 & 1 \\
$L_{9}$ & $\underline{1}\oplus\underline{1}\oplus\underline{1}$ &
$\underline{1}\oplus\underline{1}\oplus\underline{1}$ & 0 & 0 \\
\end{tabular}
\end{table}

\begin{table}
\caption{\label{quarktab} Possible Quark $SU(2)_{H}$ Representation
Assignments}
\begin{tabular}{cccccc}
 & $q_{L}$ & $u_{R}$ & $d_{R}$ & Gauge Anomaly & Global Anomaly \\
 & & & & Contribution & (\# of doublets) \\ \tableline
$Q_{1}$ & \underline{3} & \underline{3} & \underline{3} & 0 & 0 \\
$Q_{2}$ & \underline{3} & $\underline{2}\oplus\underline{1}$ &
\underline{3} & 12 & 3 \\
$Q_{3}$ & \underline{3} &
\underline{3} & $\underline{2}\oplus\underline{1}$ & $-6$ & 3 \\
$Q_{4}$ & $\underline{2}\oplus\underline{1}$ &
\underline{3} & \underline{3} & $-6$ & 6 \\
$Q_{5}$ & \underline{3} & $\underline{2}\oplus\underline{1}$ &
$\underline{2}\oplus\underline{1}$ & 6 & 6 \\
$Q_{6}$ & $\underline{2}\oplus\underline{1}$ &
$\underline{2}\oplus\underline{1}$ & \underline{3} & 6 & 9 \\
$Q_{7}$ & $\underline{2}\oplus\underline{1}$ & \underline{3} &
$\underline{2}\oplus\underline{1}$ & $-12$ & 9 \\
$Q_{8}$ & $\underline{2}\oplus\underline{1}$ &
$\underline{2}\oplus\underline{1}$ &
$\underline{2}\oplus\underline{1}$ & 0 & 12 \\
$Q_{9}$ & \underline{3} &
$\underline{1}\oplus\underline{1}\oplus\underline{1}$ & \underline{3} &
16 & 0 \\
$Q_{10}$ & \underline{3} & \underline{3} &
$\underline{1}\oplus\underline{1}\oplus\underline{1}$ & $-8$ & 0 \\
$Q_{11}$ & $\underline{1}\oplus\underline{1}\oplus\underline{1}$ &
\underline{3} & \underline{3} & $-8$ & 0 \\
$Q_{12}$ & \underline{3} &
$\underline{1}\oplus\underline{1}\oplus\underline{1}$ &
$\underline{1}\oplus\underline{1}\oplus\underline{1}$ & 8 & 0 \\
$Q_{13}$ & $\underline{1}\oplus\underline{1}\oplus\underline{1}$ &
$\underline{1}\oplus\underline{1}\oplus\underline{1}$ &
\underline{3} & 8 & 0 \\
$Q_{14}$ & $\underline{1}\oplus\underline{1}\oplus\underline{1}$ &
\underline{3} & $\underline{1}\oplus\underline{1}\oplus\underline{1}$ &
$-16$ & 0 \\
\end{tabular} \end{table}
\begin{table} \begin{tabular} {cccccc}
$Q_{15}$ & $\underline{2}\oplus\underline{1}$ &
$\underline{1}\oplus\underline{1}\oplus\underline{1}$ &
$\underline{2}\oplus\underline{1}$ & 4 & 9 \\
$Q_{16}$ & $\underline{2}\oplus\underline{1}$ &
$\underline{2}\oplus\underline{1}$ &
$\underline{1}\oplus\underline{1}\oplus\underline{1}$ & $-2$ & 9 \\
$Q_{17}$ & $\underline{1}\oplus\underline{1}\oplus\underline{1}$ &
$\underline{2}\oplus\underline{1}$ &
$\underline{2}\oplus\underline{1}$ & $-2$ & 6 \\
$Q_{18}$ & $\underline{2}\oplus\underline{1}$ &
$\underline{1}\oplus\underline{1}\oplus\underline{1}$ &
$\underline{1}\oplus\underline{1}\oplus\underline{1}$ & 2 & 6 \\
$Q_{19}$ & $\underline{1}\oplus\underline{1}\oplus\underline{1}$ &
$\underline{1}\oplus\underline{1}\oplus\underline{1}$ &
$\underline{2}\oplus\underline{1}$ & 2 & 3 \\
$Q_{20}$ & $\underline{1}\oplus\underline{1}\oplus\underline{1}$ &
$\underline{2}\oplus\underline{1}$ &
$\underline{1}\oplus\underline{1}\oplus\underline{1}$ & $-4$ & 3 \\
$Q_{21}$ & $\underline{1}\oplus\underline{1}\oplus\underline{1}$ &
$\underline{1}\oplus\underline{1}\oplus\underline{1}$ &
$\underline{1}\oplus\underline{1}\oplus\underline{1}$ & 0 & 0 \\
$Q_{22}$ & \underline{3} & $\underline{2}\oplus\underline{1}$ &
$\underline{1}\oplus\underline{1}\oplus\underline{1}$ & 4 & 3 \\
$Q_{23}$ & \underline{3} &
$\underline{1}\oplus\underline{1}\oplus\underline{1}$ &
$\underline{2}\oplus\underline{1}$ & 10 & 3 \\
$Q_{24}$ & $\underline{2}\oplus\underline{1}$ & \underline{3} &
$\underline{1}\oplus\underline{1}\oplus\underline{1}$ & $-14$ & 6 \\
$Q_{25}$ & $\underline{2}\oplus\underline{1}$ &
$\underline{1}\oplus\underline{1}\oplus\underline{1}$ &
\underline{3} & 10 & 6 \\
$Q_{26}$ & $\underline{1}\oplus\underline{1}\oplus\underline{1}$ &
\underline{3} & $\underline{2}\oplus\underline{1}$ & $-14$ & 3 \\
$Q_{27}$ & $\underline{1}\oplus\underline{1}\oplus\underline{1}$ &
$\underline{2}\oplus\underline{1}$ &\underline{3} & 4 & 3 \\
\end{tabular}
\end{table}

\begin{table}
\caption{\label{sumtab} Summary of Tree Level Results.}
\begin{tabular}{c|c|c|c}
Models & Higgs & Mass Relations & KM Matrix \\ \tableline
$L_{8} + Q_{16}$ & \mbox{$\phi \sim \underline{1}$}\ & masses for
u, c, t (two being equal); b; $\tau$ &
$U_{KM} = I_{3\times3}$ \\ \cline{2-4}
 & \mbox{$\chi \sim \underline{2}$}\ & masses for c, t; b; $\tau$
 & 1--3 mixing
angle \\ \cline{2-4}
 & \mbox{$\Delta \sim \underline{3}$}\ & masses for c, t &
unphysical \\ \hline
$L_{7} + Q_{18}$ & \mbox{$\phi \sim \underline{1}$}\ & masses for t; b;
$\tau$ & $U_{KM} = I_{3\times3}$ \\ \cline{2-4}
 & \mbox{$\chi \sim \underline{2}$}\ & masses for t; b; $\tau$ & 2--3
mixing angle \\ \hline
$L_{6} + Q_{11}$ & \mbox{$\Delta \sim \underline{3}$}\ & masses for
t; b; $\tau$ & 1--3 and 2--3 mixing angles \\ \hline
$L_{6} + Q_{10}$ & \mbox{$\phi \sim \underline{1}$}\ & $m_{u} = m_{c} =
m_{t}$ & unphysical \\ \cline{2-4}
 & \mbox{$\Delta \sim \underline{3}$}\ & masses for c, t; b; $\tau$
 & zero diagonal entries \\ \cline{2-4}
 & \mbox{$\psi \sim \underline{5}$}\ & masses for u, c, t &
unphysical \\ \hline
$L_{5} + Q_{13}$ & \mbox{$\phi \sim \underline{1}$}\ & masses for
u, c, t & unphysical \\ \cline{2-4}
 & \mbox{$\Delta \sim \underline{3}$}\ & masses for b; $\tau$ &
unphysical \\ \hline
$L_{5} + Q_{12}$ & \mbox{$\Delta \sim \underline{3}$}\ & masses for t;
b; $\tau$ & 1--3 and 2--3 mixing angles \\ \hline
$L_{3} + Q_{4}$ & \mbox{$\chi \sim \underline{2}$}\ & $m_{c} =
\frac{m_{t}}{\sqrt{2}},\; m_{s} = \frac{m_{b}}{\sqrt{2}},\; m_{\mu} =
\frac{m_{\tau}}{\sqrt{2}}$ & 2--3 mixing angle \\ \cline{2-4}
 & \mbox{$\Delta \sim \underline{3}$}\ & masses for t; b; $\tau$ &
$U_{KM} = I_{3\times3}$ \\ \cline{2-4}
 & \mbox{$\zeta \sim \underline{4}$}\ & masses for c, t; s, b; $\mu$,
$\tau$ & 2--3 mixing angle \\ \hline
$L_{2} + Q_{6}$ & \mbox{$\phi \sim \underline{1}$}\ & masses for u, c,
t --- two masses equal & unphysical \\ \cline{2-4}
 & \mbox{$\chi \sim \underline{2}$}\
 & $m_{s} = \frac{m_{b}}{\sqrt{2}},\;
m_{\mu} = \frac{m_{\tau}}{\sqrt{2}}$, masses for c, t &
zero diagonal entries \\ \cline{2-4}
 & \mbox{$\Delta \sim \underline{3}$}\ & masses for c, t; b; $\tau$
 & 2--3 mixing angle \\ \cline{2-4}
 & \mbox{$\zeta \sim \underline{4}$}\ & masses for s, b; $\mu$, $\tau$
 & unphysical \\
\end{tabular} \end{table}
\begin{table} \begin{tabular} {c|c|c|c}
$L_{1} + Q_{8}$ & \mbox{$\phi \sim \underline{1}$}\ & all massive,
many equal masses & $U_{KM} = I_{3\times3}$ \\ \cline{2-4}
 & \mbox{$\chi \sim \underline{2}$}\ & masses for c, t; s, b &
Cabibbo angle \\ \cline{2-4}
 & \mbox{$\Delta \sim \underline{3}$}\ & masses for c, t; s, b; $\tau$
 & $U_{KM} = I_{3\times3}$ \\ \cline{2-4}
 & \mbox{$\psi \sim \underline{5}$}\ & masses for e, $\mu$, $\tau$ &
unphysical \\ \hline
$L_{1} + Q_{1}$ & \mbox{$\phi \sim \underline{1}$}\ & three equal
masses for each sector & $U_{KM} = I_{3\times3}$ \\ \cline{2-4}
 & \mbox{$\Delta \sim \underline{3}$}\ & masses for t; b; $\tau$
 & 1--3 and 2--3 mixing angles \\ \cline{2-4}
 & \mbox{$\psi \sim \underline{5}$}\ & all massive
& all entries non-zero \\ \hline
$L_{9} + Q_{8}$ & \mbox{$\phi \sim \underline{1}$}\ & all massive,
some equal & $U_{KM} = I_{3\times3}$ \\ \cline{2-4}
 & \mbox{$\chi \sim \underline{2}$}\ & masses for c, t; s, b &
Cabibbo angle \\ \cline{2-4}
 & \mbox{$\Delta \sim \underline{3}$}\ & masses for c, t; s, b &
$U_{KM} = I_{3\times3}$ \\ \hline
$L_{9} + Q_{1}$ & \mbox{$\phi \sim \underline{1}$}\ & all massive,
many equal masses & $U_{KM} = I_{3\times3}$ \\ \cline{2-4}
 & \mbox{$\Delta \sim \underline{3}$}\ & masses for t; b & 1--3 and
2--3 mixing angles \\ \cline{2-4}
 & \mbox{$\psi \sim \underline{5}$}\ & masses for all quarks & all
entries non-zero \\ \hline
$L_{1} + Q_{21}$ & \mbox{$\phi \sim \underline{1}$}\ & all massive,
$m_{e} = m_{\mu} = m_{\tau}$ & all entries non-zero \\ \cline{2-4}
 & \mbox{$\Delta \sim \underline{3}$}\ & mass for $\tau$ only &
unphysical \\ \cline{2-4}
 & \mbox{$\psi \sim \underline{5}$}\ & masses for e, $\mu$, $\tau$ &
unphysical \\
\end{tabular}
\end{table}

\begin{table}
\squeezetable
\caption{\label{rhnsummary} Summary of Results for Models with RHNs.}
\begin{tabular}{c|c|c|c|c}
Models & Higgs & Mass Relations & KM Matrix & Lepton Matrix \\
\tableline
$L_{8} + Q_{17} + N_{2}$ & \mbox{$\phi \sim \underline{1}$}\ & masses
for t; b; $\tau$; $\nu_{3}$ & 1--3 and 2--3 & 1--3 and 2--3 \\ & &
 & mixing angles & mixing angles\\ \cline{2-5}
 & \mbox{$\chi \sim \underline{2}$}\ & masses for t; b; $\tau$;
$\nu_{3}$ & 1--3 and 2--3 &
1--3 and 2--3 \\ & & & mixing angles & mixing angles\\ \hline
$L_{8} + Q_{16} + N_{1}$ & \mbox{$\phi \sim \underline{1}$}\ & masses
for u, c, t & $U_{KM} = I_{3\times3}$ & all entries \\
& & (two being equal); b; $\tau$; all $\nu$'s & & non-zero \\
\cline{2-5}
 & \mbox{$\chi \sim \underline{2}$}\ & masses for c, t; b; $\tau$
 & 1--3 mixing angle & unphysical \\ \cline{2-5}
 & \mbox{$\Delta \sim \underline{3}$}\ & masses for c, t & unphysical
 & unphysical \\ \hline $L_{8} + Q_{16} + N_{3}$
 & \mbox{$\phi \sim \underline{1}$}\ & masses for
u, c, t & $U_{KM} = I_{3\times3}$ & unphysical \\
& & (two being equal); b; $\tau$ & & \\ \cline{2-5}
 & \mbox{$\chi \sim \underline{2}$}\ & masses for c, t; b; $\tau$
 & 1--3 mixing angle & unphysical \\ \cline{2-5}
 & \mbox{$\Delta \sim \underline{3}$}\ & masses for c, t; $\nu_{3}$
 & unphysical & unphysical \\ \hline
$L_{7} + Q_{19} + N_{2}$ & \mbox{$\phi \sim \underline{1}$}\ & masses
for u, c, t; & all entries
& $U_{e\nu} = I_{3\times3}$ \\ & & b; $\tau$; $\nu_{1} = \nu_{2} =
\nu_{3}$ & non-zero & \\ \cline{2-5}
 & \mbox{$\chi \sim \underline{2}$}\ & masses for b; $\tau$;
$\nu_{2}$, $\nu_{3}$ & unphysical & 1--3 mixing angle \\ \cline{2-5}
 & \mbox{$\Delta \sim \underline{3}$}\ & masses for $\nu_{2}$,
$\nu_{3}$ & unphysical & unphysical \\ \hline
$L_{7} + Q_{18} + N_{1}$ & \mbox{$\phi \sim \underline{1}$}\ & masses
for t; b; $\tau$; $\nu_{3}$ & $U_{KM} = I_{3\times3}$ & $U_{e\nu} =
I_{3\times3}$ \\ \cline{2-5}
 & \mbox{$\chi \sim \underline{2}$}\ & masses for t; b; $\tau$;
$\nu_{3}$ & 2--3 mixing angle
 & 1--3 or 2--3 \\ & & & & mixing angle \\ \hline
$L_{7} + Q_{18} + N_{3}$ & \mbox{$\phi \sim \underline{1}$}\ & masses
for t; b; $\tau$ & $U_{KM} = I_{3\times3}$ & unphysical \\ \cline{2-5}
 & \mbox{$\chi \sim \underline{2}$}\ & masses for t; b; $\tau$; & 2--3
mixing angle & 2--3 mixing angle \\ & & $m_{\nu_{2}} =
\frac{m_{\nu_{3}}}{\sqrt{2}}$ & & \\ \hline
$L_{6} + Q_{11} + N_{1}$ & \mbox{$\phi \sim \underline{1}$}\ & all
$\nu$'s massive & unphysical & unphysical \\ \cline{2-5}
 & \mbox{$\Delta \sim \underline{3}$}\ & masses for t; b; $\tau$
& 1--3 and 2--3 & unphysical \\ & & & mixing angles & \\ \hline
$L_{6} + Q_{11} + N_{3}$ & \mbox{$\Delta \sim \underline{3}$}\
 & masses for t; b;
$\tau$; $\nu_{3}$ & 1--3 and 2--3 & 1--3 and 2--3 \\ & &
& mixing angles & mixing angles \\ \hline
$L_{6} + Q_{10} + N_{1}$ & \mbox{$\phi \sim \underline{1}$}\
 & $m_{u} = m_{c} =
m_{t}$, & unphysical & unphysical \\ & & all neutrinos massive & &
\\ \cline{2-5}
 & \mbox{$\Delta \sim \underline{3}$}\ & masses for c, t; b; $\tau$
 & zero diagonal & unphysical \\ & & & entries & \\ \cline{2-5}
 & \mbox{$\psi \sim \underline{5}$}\ & masses for u, c, t &
unphysical & unphysical \\
\end{tabular} \end{table}
\begin{table} \squeezetable \begin{tabular} {c|c|c|c|c}
$L_{6} + Q_{10} + N_{3}$ & \mbox{$\phi \sim \underline{1}$}\
 & $m_{u} = m_{c} = m_{t}$ & unphysical & unphysical \\ \cline{2-5}
 & \mbox{$\Delta \sim \underline{3}$}\ & masses for c, t; b; $\tau$;
$\nu_{3}$ & zero diagonal & 1--3 and 2--3 \\ & & & entries & mixing
angles \\ \cline{2-5}
 & \mbox{$\psi \sim \underline{5}$}\ & masses for u, c, t &
unphysical & unphysical \\ \hline
$L_{5} + Q_{13} + N_{1}$ & \mbox{$\phi \sim \underline{1}$}\ & masses
for u, c, t & unphysical & unphysical \\ \cline{2-5}
 & \mbox{$\Delta \sim \underline{3}$}\ & masses for b; $\tau$;
$\nu_{3}$ & unphysical
 & 1--3 and 2--3 \\ & & & & mixing angles \\ \hline
$L_{5} + Q_{13} + N_{3}$ & \mbox{$\phi \sim \underline{1}$}\ & masses
for u, c, t; & unphysical & unphysical \\ & & 3 equal $\nu$
masses & & \\ \cline{2-5}
 & \mbox{$\Delta \sim \underline{3}$}\ & masses for b; $\tau$;
$\nu_{2}$, $\nu_{3}$ &
unphysical & zero diagonal \\ & & & & entries \\ \hline
$L_{5} + Q_{12} + N_{1}$ & \mbox{$\Delta \sim \underline{3}$}\
 & masses for t; b;
$\tau$; $\nu_{3}$ & 1--3 and 2--3 & 1--3 and 2--3 \\
& & & mixing angles & mixing angles \\ \hline
$L_{5} + Q_{12} + N_{3}$ & \mbox{$\phi \sim \underline{1}$}\ & 3 equal
$\nu$ masses & unphysical & unphysical \\ \cline{2-5}
 & \mbox{$\Delta \sim \underline{3}$}\ & masses for t; b; $\tau$;
$\nu_{2}$, $\nu_{3}$ &
1--3 and 2--3 & zero diagonal \\ & & & mixing angles & entries
\\ \cline{2-5}
 & \mbox{$\psi \sim \underline{5}$}\ & 3 $\nu$ masses & unphysical
 & unphysical \\ \hline
$L_{4} + Q_{8} + N_{2}$ & \mbox{$\phi \sim \underline{1}$}\ & all
massive, & $U_{KM} =
I_{3\times3}$ & $U_{e\nu} = I_{3\times3}$ \\ & & many equal masses & &
\\ \cline{2-5}
 & \mbox{$\chi \sim \underline{2}$}\ & masses for c, t; s, b; $\mu$,
$\tau$; $\nu_{2}$,
$\nu_{3}$ & Cabibbo angle & 1--2 or 1--3 \\ & & & & mixing angle \\
\cline{2-5}
 & \mbox{$\Delta \sim \underline{3}$}\ & masses for c, t; s, b; $\mu$,
$\tau$; $\nu_{2}$,
$\nu_{3}$ & $U_{KM} = I_{3\times3}$ & $U_{e\nu} = I_{3\times3}$ \\
\hline
$L_{4} + Q_{1} + N_{2}$ & \mbox{$\phi \sim \underline{1}$}\ & all
massive, & $U_{KM} =
I_{3\times3}$ & $U_{e\nu} = I_{3\times3}$ \\ & & many equal masses & &
\\ \cline{2-5}
 & \mbox{$\chi \sim \underline{2}$}\ & masses for $\mu$, $\tau$;
$\nu_{2}$, $\nu_{3}$ &
unphysical & 1--2 or 1--3 \\ & & & & mixing angle \\ \cline{2-5}
 & \mbox{$\Delta \sim \underline{3}$}\ & masses for t; b; $\mu$,
$\tau$; $\nu_{2}$, $\nu_{3}$ &
1--3 and 2--3 & $U_{e\nu} = I_{3\times3}$ \\ & & & mixing angles &
\\ \cline{2-5}
 & \mbox{$\psi \sim \underline{5}$}\ & masses for u, c, t; d, s, b
 & all entries & unphysical
\\ & & & non-zero & \\ \hline
$L_{3} + Q_{4} + N_{1}$ & \mbox{$\phi \sim \underline{1}$}\ & mass for
$\nu_{3}$ only & unphysical & unphysical \\ \cline{2-5}
 & \mbox{$\chi \sim \underline{2}$}\
 & $m_{c} = \frac{m_{t}}{\sqrt{2}};\;
m_{s} = \frac{m_{b}}{\sqrt{2}}$; & 2--3 mixing angle & 1--3 or 2--3
\\ & & $m_{\mu} = \frac{m_{\tau}}{\sqrt{2}}$;
mass for $\nu_{3}$ & & mixing angle \\ \cline{2-5}
 & \mbox{$\Delta \sim \underline{3}$}\ & masses for t; b; $\tau$ &
$U_{KM} = I_{3\times3}$ & unphysical \\ \cline{2-5}
 & \mbox{$\zeta \sim \underline{4}$}\ & masses for c, t; s, b; $\mu$,
$\tau$ & 2--3 mixing angle & unphysical \\
\end{tabular} \end{table}
\begin{table} \squeezetable \begin{tabular}{c|c|c|c|c}
$L_{3} + Q_{4} + N_{3}$ & \mbox{$\chi \sim \underline{2}$}\ & $m_{c} =
\frac{m_{t}}{\sqrt{2}};\;
m_{s} = \frac{m_{b}}{\sqrt{2}}$;
& 2--3 mixing angle & 2--3 mixing angle \\ & & $m_{\mu} =
\frac{m_{\tau}}{\sqrt{2}};\; m_{\nu_{2}} =
\frac{m_{\nu_{3}}}{\sqrt{2}}$ & & \\ \cline{2-5}
 & \mbox{$\Delta \sim \underline{3}$}\ & masses for t; b; $\tau$;
$\nu_{3}$ & $U_{KM} = I_{3\times3}$
 & $U_{e\nu} = I_{3\times3}$ \\ \cline{2-5}
 & \mbox{$\zeta \sim \underline{4}$}\ & masses for c, t; s, b; $\mu$,
$\tau$; $\nu_{2}$,
$\nu_{3}$ & 2--3 mixing angle & 2--3 mixing angle \\ \hline
$L_{3} + Q_{3} + N_{2}$ & \mbox{$\phi \sim \underline{1}$}\
 & $m_{u} = m_{c} = m_{t}$; &
unphysical & unphysical \\ & & 3 $\nu$ masses --- two equal & & \\
\cline{2-5} & \mbox{$\chi \sim \underline{2}$}\
 & $m_{s} = \frac{m_{b}}{\sqrt{2}};\; m_{\mu} =
\frac{m_{\tau}}{\sqrt{2}}$; & unphysical & zero diagonal
\\ & & masses for $\nu_{2}$, $\nu_{3}$ & & entries \\ \cline{2-5}
 & \mbox{$\Delta \sim \underline{3}$}\ & masses for t; b; $\tau$;
$\nu_{2}$, $\nu_{3}$ & 1--3 and
2--3 & all entries \\ & & & mixing angles & non-zero \\ \cline{2-5}
 & \mbox{$\zeta \sim \underline{4}$}\ & masses for s, b; $\mu$, $\tau$
 & unphysical & unphysical
\\ \cline{2-5}
 & \mbox{$\psi \sim \underline{5}$}\ & masses for u, c, t & unphysical
 & unphysical \\ \hline
$L_{2} + Q_{6} + N_{1}$ & \mbox{$\phi \sim \underline{1}$}\ & masses
for u, c, t --- & unphysical
& unphysical \\ & & two masses equal & & \\ \cline{2-5}
 & \mbox{$\chi \sim \underline{2}$}\
 & $m_{s} = \frac{m_{b}}{\sqrt{2}},\;
m_{\mu} = \frac{m_{\tau}}{\sqrt{2}}$; &
zero diagonal & unphysical \\ & & masses for c, t & entries &
\\ \cline{2-5}
 & \mbox{$\Delta \sim \underline{3}$}\ & masses for c, t; b; $\tau$;
$\nu_{3}$ & 2--3
mixing angle & 1--3 and 2--3 \\ & & & & mixing angles \\ \cline{2-5}
 & \mbox{$\zeta \sim \underline{4}$}\ & masses for s, b; $\mu$, $\tau$
 & unphysical & unphysical \\ \hline
$L_{2} + Q_{6} + N_{3}$ & \mbox{$\phi \sim \underline{1}$}\ & masses
for u, c, t (two equal); & unphysical & unphysical \\ & & 3 equal $\nu$
masses & & \\ \cline{2-5}
 & \mbox{$\chi \sim \underline{2}$}\
 & $m_{s} = \frac{m_{b}}{\sqrt{2}},\;
m_{\mu} = \frac{m_{\tau}}{\sqrt{2}}$; &
zero diagonal & unphysical \\ & & masses for c, t & entries &
\\ \cline{2-5}
 & \mbox{$\Delta \sim \underline{3}$}\ & masses for c, t; b; $\tau$;
$\nu_{3}$ & 2--3
mixing angle & 1--3 and 2--3 \\ & & & & mixing angles \\ \cline{2-5}
 & \mbox{$\zeta \sim \underline{4}$}\ & masses for s, b; $\mu$, $\tau$
 & unphysical & unphysical \\ \cline{2-5}
 & \mbox{$\psi \sim \underline{5}$}\ & 3 $\nu$ masses & unphysical
 & unphysical \\ \hline
$L_{2} + Q_{5} + N_{2}$ & \mbox{$\chi \sim \underline{2}$}\ & $m_{c} =
\frac{m_{t}}{\sqrt{2}};\;
m_{s} = \frac{m_{b}}{\sqrt{2}}$; & zero diagonal & zero
diagonal \\ & & $m_{\mu} = \frac{m_{\tau}}{\sqrt{2}};\;
m_{\nu_{2}} = \frac{m_{\nu_{3}}}{\sqrt{2}}$ & entries & entries
\\ \cline{2-5}
 & \mbox{$\Delta \sim \underline{3}$}\ & masses for t; b; $\tau$;
$\nu_{3}$ & 1--3 and 2--3 & 1--3
and 2--3 \\ & & & mixing angles & mixing angles \\ \cline{2-5}
 & \mbox{$\zeta \sim \underline{4}$}\ & masses for c, t; s, b; $\mu$,
$\tau$; $\nu_{2}$, $\nu_{3}$
& all entries & all entries \\ & & & non-zero & non-zero \\
\end{tabular} \end{table}
\begin{table} \squeezetable \begin{tabular}{c|c|c|c|c}
$L_{1} + Q_{8} + N_{1}$ & \mbox{$\phi \sim \underline{1}$}\ & all
except $\nu$'s massive, &
$U_{KM} = I_{3\times3}$ & unphysical \\ & & many equal masses & &
\\ \cline{2-5}
 & \mbox{$\chi \sim \underline{2}$}\ & masses for c, t; s, b &
Cabibbo angle & unphysical \\ \cline{2-5}
 & \mbox{$\Delta \sim \underline{3}$}\ & masses for c, t; s, b;
$\tau$; $\nu_{3}$ &
$U_{KM} = I_{3\times3}$ & 1--3 and 2--3 \\ & & & & mixing angles
\\ \cline{2-5}
 & \mbox{$\psi \sim \underline{5}$}\ & masses for e, $\mu$, $\tau$ &
unphysical & unphysical \\ \hline
$L_{1} + Q_{8} + N_{3}$ & \mbox{$\phi \sim \underline{1}$}\ & all
massive, many equal
masses & $U_{KM} = I_{3\times3}$ & $U_{e\nu} = I_{3\times3}$
\\ \cline{2-5}
 & \mbox{$\chi \sim \underline{2}$}\ & masses for c, t; s, b &
Cabibbo angle & unphysical \\ \cline{2-5}
 & \mbox{$\Delta \sim \underline{3}$}\ & masses for c, t; s, b;
$\tau$; $\nu_{3}$ &
$U_{KM} = I_{3\times3}$ & 1--3 and 2--3 \\ & & & & mixing angles
\\ \cline{2-5}
 & \mbox{$\psi \sim \underline{5}$}\ & masses for e, $\mu$, $\tau$;
3 $\nu$'s & unphysical & 1--3 and 2--3 \\ & & &
 & mixing angles \\ \hline
$L_{1} + Q_{1} + N_{1}$ & \mbox{$\phi \sim \underline{1}$}\ & three
equal masses for &
$U_{KM} = I_{3\times3}$ & unphysical \\ & & quarks and charged
leptons & & \\ \cline{2-5}
 & \mbox{$\Delta \sim \underline{3}$}\ & masses for t; b; $\tau$;
$\nu_{3}$ & 1--3 and
2--3 & 1--3 and 2--3 \\ & & & mixing angles & mixing angles
\\ \cline{2-5}
 & \mbox{$\psi \sim \underline{5}$}\ & all massive except $\nu$'s
& all entries & unphysical \\ & & & non-zero & \\ \hline
$L_{1} + Q_{1} + N_{3}$ & \mbox{$\phi \sim \underline{1}$}\ & three
equal masses for
each sector & $U_{KM} = I_{3\times3}$ & $U_{e\nu} = I_{3\times3}$
\\ \cline{2-5}
 & \mbox{$\Delta \sim \underline{3}$}\ & masses for t; b; $\tau$;
$\nu_{3}$ & 1--3 and
2--3 & 1--3 and 2--3 \\ & & & mixing angles & mixing angles
\\ \cline{2-5}
 & \mbox{$\psi \sim \underline{5}$}\ & all massive
& all entries & all entries \\ & & & non-zero & non-zero \\ \hline
$L_{9} + Q_{8} + N_{1}$ & \mbox{$\phi \sim \underline{1}$}\ & all
massive, some equal &
$U_{KM} = I_{3\times3}$ & all entries \\ & & & & non-zero \\
\cline{2-5}
 & \mbox{$\chi \sim \underline{2}$}\ & masses for c, t; s, b &
Cabibbo angle & unphysical \\ \cline{2-5}
 & \mbox{$\Delta \sim \underline{3}$}\ & masses for c, t; s, b &
$U_{KM} = I_{3\times3}$ & unphysical \\ \hline
$L_{9} + Q_{8} + N_{3}$ & \mbox{$\phi \sim \underline{1}$}\ & all
massive except $\nu$'s, &
$U_{KM} = I_{3\times3}$ & unphysical \\ & & some equal masses & &
\\ \cline{2-5}
 & \mbox{$\chi \sim \underline{2}$}\ & masses for c, t; s, b &
Cabibbo angle & unphysical \\ \cline{2-5}
 & \mbox{$\Delta \sim \underline{3}$}\ & masses for c, t; s, b;
$\nu_{3}$ & $U_{KM} = I_{3\times3}$ & unphysical \\
\end{tabular} \end{table}
\begin{table} \squeezetable \begin{tabular}{c|c|c|c|c}
$L_{9} + Q_{1} + N_{1}$ & \mbox{$\phi \sim \underline{1}$}\ & all
massive, & $U_{KM} = I_{3\times3}$ & all entries \\ & & many equal
masses & & non-zero \\ \cline{2-5}
 & \mbox{$\Delta \sim \underline{3}$}\ & masses for t; b & 1--3 and
2--3 & unphysical \\ & & & mixing angles & \\ \cline{2-5}
 & \mbox{$\psi \sim \underline{5}$}\ & masses for all quarks & all
entries & unphysical \\ & & & non-zero & \\ \hline
$L_{9} + Q_{1} + N_{3}$ & \mbox{$\phi \sim \underline{1}$}\ & all
massive except $\nu$'s,
& $U_{KM} = I_{3\times3}$ & unphysical \\ & & many equal masses & &
\\ \cline{2-5}
 & \mbox{$\Delta \sim \underline{3}$}\ & masses for t; b; $\nu_{3}$
 & 1--3 and 2--3 & unphysical \\ & & & mixing angles & \\ \cline{2-5}
 & \mbox{$\psi \sim \underline{5}$}\ & masses for all quarks & all
entries & unphysical \\ & & & non-zero & \\ \hline
$L_{4} + Q_{21} + N_{2}$ & \mbox{$\phi \sim \underline{1}$}\ & all
massive, some & all entries &
$U_{e\nu} = I_{3\times3}$ \\ & & equal masses & non-zero &
\\ \cline{2-5}
 & \mbox{$\chi \sim \underline{2}$}\ & masses for $\mu$, $\tau$;
$\nu_{2}$, $\nu_{3}$ &
unphysical & 1--2 or 1--3 \\ & & & & mixing angle \\ \cline{2-5}
 & \mbox{$\Delta \sim \underline{3}$}\ & masses for $\mu$, $\tau$;
$\nu_{2}$, $\nu_{3}$ & unphysical & $U_{e\nu} = I_{3\times3}$ \\ \hline
$L_{1} + Q_{21} + N_{1}$ & \mbox{$\phi \sim \underline{1}$}\
 & $m_{e} = m_{\mu} = m_{\tau}$,
& all entries & unphysical \\ & & masses for all quarks & non-zero &
\\ \cline{2-5}
 & \mbox{$\Delta \sim \underline{3}$}\ & masses for $\tau$; $\nu_{3}$
 & unphysical & 1--3 and 2--3 \\ & & & & mixing angles \\ \cline{2-5}
 & \mbox{$\psi \sim \underline{5}$}\ & masses for e, $\mu$, $\tau$ &
unphysical & unphysical \\ \hline
$L_{1} + Q_{21} + N_{3}$ & \mbox{$\phi \sim \underline{1}$}\ & all
massive, 3 equal & all entries & $U_{e\nu} = I_{3\times3}$ \\ &
 & masses for lepton sectors & non-zero & \\ \cline{2-5}
 & \mbox{$\Delta \sim \underline{3}$}\ & masses for $\tau$; $\nu_{3}$
 & unphysical & 1--3 and 2--3 \\ & & & & mixing angles \\ \cline{2-5}
 & \mbox{$\psi \sim \underline{5}$}\ & masses for all leptons &
unphysical & all entries \\ & & & & non-zero \\
\end{tabular}
\end{table}

\begin{table}
\caption{\label{nu-bounds} Bounds on Neutrino masses from Accelerator\\
and Double-Beta Decay Measurements\protect\cite{DC&PL,PDG7}.}
\begin{tabular}{l}
Accelerator Limits \\ \tableline
$\nu_{e}$ \qquad \qquad $<17$eV \\ \tableline
$\nu_{\mu}$ \qquad \qquad $<0.27$MeV \\ \tableline
$\nu_{\tau}$ \qquad \qquad $<35$MeV \\ \tableline
Double-Beta Decay Limit \\ \tableline
$\langle m_{\nu_{e}} \rangle =
\sum_{i} \xi_{i}U_{ei}^{2}m_{i} < (2 \pm 1)\;$ eV
\end{tabular}
\end{table}

\end{document}